\documentclass[%
reprint,
 amsmath,amssymb,
 aps,
]{revtex4-2}

\usepackage{aas_macros}
\usepackage{graphicx}
\usepackage{dcolumn}
\usepackage{bm}
\usepackage{hyperref}
\usepackage{color}
\usepackage{natbib}
\usepackage{CJK}
\usepackage[normalem]{ulem}

\usepackage[english]{babel}
\usepackage{xspace}
\usepackage{ulem}

\newcommand{\rvw}[1]{#1}

\newcommand{\harm}{\xspace{\sc Harm3d}\xspace}

\begin{document}
\begin{CJK*}{UTF8}{gbsn} 
\preprint{APS/123-QED}

\title{General Relativistic magneto-hydrodynamical simulations of accretion flows through traversable wormholes}

\author{Luciano Combi$^{1,2}$, Huan Yang$^{1,2,3}$, Eduardo Gutierrez$^{4,5}$, Scott C. Noble$^{6}$, Gustavo E. Romero$^{7,8}$, Manuela Campanelli$^{9}$}
\affiliation{
 $^1$Perimeter Institute for Theoretical Physics, Waterloo, Ontario, Canada, N2L 2Y5\\
 $^2$Department of Physics, University of Guelph, Guelph, Ontario, Canada, N1G 2W1\\ 
 $^3$ Department of Astronomy, Tsinghua University, Beijing 100084, China \\ 
 $^{4}$Institute for Gravitation and the Cosmos, The Pennsylvania State University, University Park PA 16802, USA\\
 $^{5}$Department of Physics, The Pennsylvania State University, University Park PA 16802, USA\\
 $^6$Gravitational Astrophysics Lab, NASA Goddard Space Flight Center, Greenbelt, MD 20771, USA\\
 $^{7}$Instituto Argentino de Radioastronom\'ia (IAR, CCT La Plata, CONICET/CIC), C.C.5, (1984) Villa Elisa, Buenos Aires, Argentina\\
 $^{8}$Facultad de Ciencias Astron\'omicas y Geof\'isicas, Universidad Nacional de La Plata, Paseo del Bosque s/n, 1900 La Plata, Buenos Aires, Argentina \\
 $^{9}$Center for Computational Relativity and Gravitation, Rochester Institute of Technology, Rochester, NY 14623, USA
 }


\begin{abstract}
\rvw{We present the first dynamical model of plasma accretion onto traversable wormholes by performing General Relativistic magneto-hydrodynamical (GRMHD) simulations of the flow on both sides of the wormhole.} We evolve the ideal MHD equations on a wormhole spacetime described by the spherically symmetric Simpson--Visser metric. The disk is initialized on one side of the wormhole and accretes onto the throat driven by the magneto-rotational instability (MRI). We show that the inflowing plasma quickly settles in the throat and forms a hot, rotating cloud. The wormhole cloud acts as an engine in which gas coming from one side accumulates at the center, dissipates energy, and powers a mildly relativistic thermal wind toward the other side. \rvw{Our novel predictions show that accreting wormholes behave very differently from black holes (BHs) in astrophysical environments. In particular, one mouth presents outflows without accretion signatures, contradicting the jet-disk symbiotic relation that holds for black holes.}

\end{abstract}

\maketitle
\end{CJK*}

\section{Introduction}
A wormhole is a topological shortcut in spacetime connecting separate regions of the Universe through a throat. These hypothetical, exotic objects provide a rich conceptual framework to understand physics in curved spacetime and its most extreme consequences \citep{morris1988wormholes}. \rvw{A stable, traversable wormhole violates the averaged null energy condition \citep{visser1995lorentzian}, which sets strong constraints on their existence \citep{lobo2017wormholes}; this usually requires non-standard physics like exotic matter with extreme equations of state \cite{visser1990wormholes} or alternative theories of gravity\citep{visser1989traversable, maldacena2021humanly, blazquez2021traversable, konoplya2022traversable,barcelo2000scalar}. Their radial and non-radial stability, on the other hand, depend on specific wormhole configurations \cite{Poisson:1995sv,Cardoso:2019rvt,Dias:2010uh,Armendariz-Picon:2002gjc,Yang:2022gic}}

\rvw{Despite the strong constraints on fundamental physics, the existence of natural wormholes cannot be completely ruled out.}
Many properties of the strong gravitational field near a BH horizon can be mimicked by the horizonless spacetime of a wormhole outside the throat \citep{damour2007wormholes}, including the spectrum of thin disks surrounding one mouth \citep{bambi2021astrophysical, harko2008electromagnetic}, and gravitational waves \citep{damour2007wormholes, cardoso2016gravitational}; in other aspects, such as lensing of light rays, wormholes can show distinctive features from normal BHs \cite{safonova2001microlensing, torres1998might}. 
All previous work on astrophysical plasmas surrounding wormholes, to our knowledge, only considered simple stationary models for matter, such as Novikov--Thorne disks surrounding one mouth, and mostly focusing on one side of the throat. 

In this work, we investigate the dynamics of magnetized plasma surrounding a wormhole by performing GRMHD simulations of an accretion disk in traversable wormhole spacetimes. To this purpose, we use the GRMHD code \harm which is capable of evolving the ideal MHD equations on an arbitrary metric with arbitrary coordinates. We use the Simpson--Visser metric as a simple model of a two-way traversable wormhole with no rotation, {assumed to be} sustained by exotic matter, and similar to a Schwarzschild BH from the outside. We analyze the plasma dynamics on both sides and across the throat for different throat sizes. We use geometrical units ($c=G=1$) and a $+2$ signature.

\section{Wormhole metric}
We are interested in analyzing accretion onto a simple traversable wormhole model. We focus on spherically-symmetric wormholes, which are completely characterized by a shape function (which defines the geometry of the throat) and a redshift factor (which defines the acceleration of stationary frames) \citep{morris1988wormholes}. For this purpose, we consider the Simpson--Visser metric \citep{simpson2019black}, a simple extension of the Schwarzschild metric:
\begin{equation}
ds^2 = - f(r) dt^2 + f(r)^{-1}dr^2 + (r^2+\ell^2) d\Omega^2,
\label{eq:metric}
\end{equation}
with $f(r):= \Big(1 - {2M}/{\sqrt{r^2 + \ell^2}} \Big)$, and $d\Omega^2 = d\theta^2 + \sin(\theta)^2 d\phi^2$. The solution parameterized with $\ell$ represents a two-way wormhole for $\ell > 2M$, a black-bounce solution for $\ell<2M$, and a Scwharzschild BH for $\ell =0$. For $\ell>0$, the solution is regular at $r=0$, and for $\ell>2M$ it can also be extended to negative values of $r$. Although the coordinate $r\in [-\infty, +\infty]$ does not represent the proper size, it maps out the entire spacetime. The wormhole solution has a traversable throat at $r=0$, which is a time-like spherical surface of proper radius $\ell$. We define mouth A as the positive entrance to the throat, $r>0$, and mouth B as the negative side with $r<0$. The spacetime on both mouths is symmetric and attractive on each side. The innermost stable orbit (ISCO) of the spacetime is located at $r_{\rm ISCO} = 6M \sqrt{1- (\ell/6M)^2}$ and the photon ring is at $r_{\rm ph} = 3M \sqrt{1- (\ell/3M)^2}$ \citep{bambhaniya2022thin}. In General Relativity, this solution requires exotic matter near the throat; we assume that the field supporting the wormhole is not dynamically affected by the accreting plasma, nor vice-versa.

\section{Simulation set-up}
We evolve the GRMHD equations using the flux-conservative code \harm \cite{noble2009direct}, which is a 3D MPI-parallelized, fully covariant, modernized version of the public code HARM \citep{gammie2003harm,  noble2006primitive}. The code allows us to simulate an accretion disk on an arbitrary spacetime with arbitrary coordinates \citep{zilhao2014dynamic}. For this application, we build a spherical grid symmetric on each side of the throat with more resolution close to the throat. The uniform-spaced numerical coordinates $(x_1,x_2,x_3)$ are mapped to physical coordinates as $r(x_1) = c_{r} \sinh(x/c_{r})$, where we fix $c_{r}= 1-\log{2}/\log{R_{\rm out}}$, with $R_{\rm out}=600$, and $\lbrace \theta(x_2), \phi(x_3) \rbrace = \lbrace x_2, x_3 \rbrace$. We use a resolution of $n_{r} \times n_{\theta} \times n_{\phi} = 460 \times 128 \times 128$, with $240$ radial cells on each side of the wormhole. The radial extension of the grid is $r \in [-600 M, 600 M]$ and we impose outflow conditions on the outer boundaries. We cover the full $(\theta, \phi)$ sphere with our grid and we impose transmissive boundary conditions on the angular boundaries \cite{armengol2022handing}.

We assume the wormhole is accreting gas {through} mouth A from a hot thick torus, as found in advection-dominated accretion flows (ADAF), either at low accretion rates, such as M87, or at high accretion rates, such as some X-ray binaries and tidal disruption events. Mouth B of the wormhole is initially set to be gas-free. The equation of state of the plasma is $u = p/(\Gamma-1)$, where $u$ is the internal energy and $p$ is the pressure. The adiabatic index is chosen as $\Gamma=4/3$, consistent with a radiation-dominated flow. We do not include any radiative cooling; all heat generated in dynamic processes, e.g. turbulence, shocks, and magnetic heating, increases the gas entropy.

The initial data consists of a disk in hydrostatic equilibrium surrounding mouth A with a power-law angular momentum profile following the procedure in Refs. \cite{chakrabarti1985natural, de2003magnetically}. We set the inner radius of the torus at $r=25 M$ and the maximum pressure of the disk at $r=35M$. Notice that at $r \approx 10M$, the difference between the wormhole metric (Eq. \ref{eq:metric}) and the Schwarzschild metric is less than $ 0.1 \%$ for $\ell = 2.1M$. In addition, we initialize a weak poloidal magnetic field with a strength set by {ensuring the ratio of the volume-averaged gas and magnetic pressures is $100$}, and small perturbations to the internal energy to {accelerate} the magneto-rotational (MRI) instability {and the plasma's development into non-axisymmetric turbulence}. In mouth B, we set all MHD variables to atmosphere values. We simulate two spacetime configurations by setting the smaller throat size $\ell = 2.1M$ and the larger throat size $\ell= 3M$. For $\ell = 2.1M$, the ISCO is at $5.2 M$ and there is a photon ring at $\approx 2.1 M$; for $\ell = 3M$, the ISCO is at $\sim 5.6M$ and the spacetime does not have a photon ring. 

We focus our discussions in the next sections on the model with the smaller throat $\ell = 2.1 M$. We will compare with the bigger throat in the last section.

\begin{figure}[ht!]
    \centering    \includegraphics[width=\columnwidth]{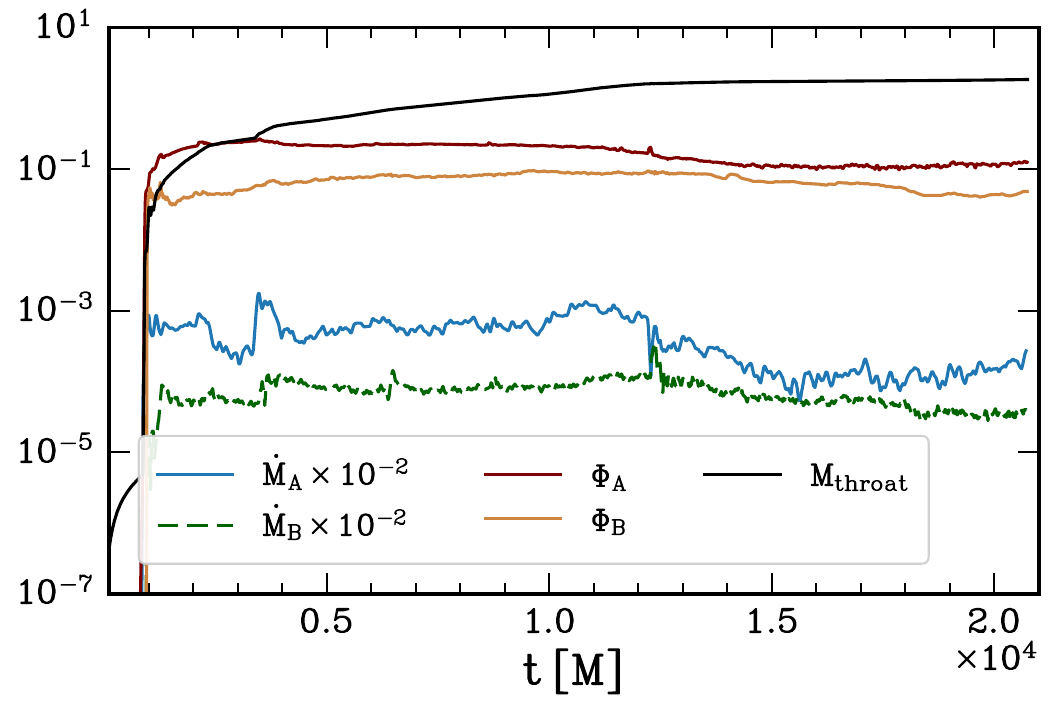}
    \caption{Mass contained in the throat ($|r|<5M$), mass flux and magnetic flux measured at $r=\pm 5M$ as a function of time for our fiducial case $\ell =2.1$. }
    \label{fig:qvst}
\end{figure}

\begin{figure}[ht!]
    \centering
    \includegraphics[width=\columnwidth]{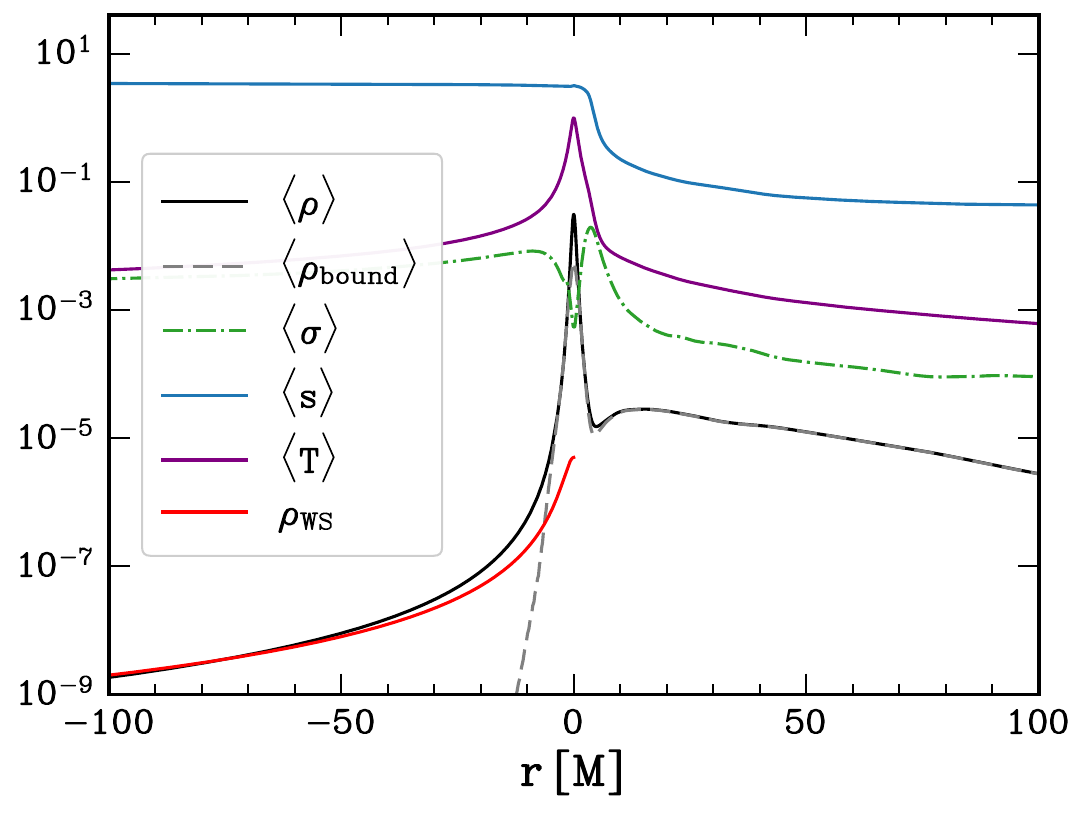}
    \caption{Surface-average properties of the accretion throat as a function of radius, averaged in time between $t \in [10, 20]\times 10^{4} M$ {for the case  $\ell=2.1M$}. We show the rest-mass density of the whole flow, $\langle \rho \rangle$, rest-mass density of the bound fluid, $\langle \rho_{\rm bound} \rangle$, where integrate only fluid elements with $-hu_t<1$), and density-weighted temperature, $\langle T \rangle$, magnetization, $\langle \sigma \rangle$ and entropy, $\langle s \rangle$. We plot the density of a spherically symmetric stationary wind, $\rho_{\rm WS} \propto r^{-2}$.}
    \label{fig:qvsr}
\end{figure}

\begin{figure*}[ht!]
    \centering
    \includegraphics[width=1.0\linewidth]{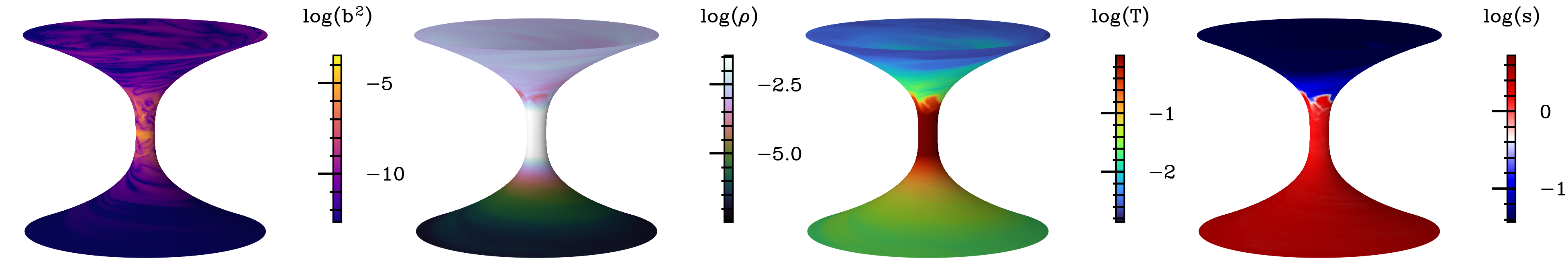}
    \includegraphics[width=0.49\linewidth]{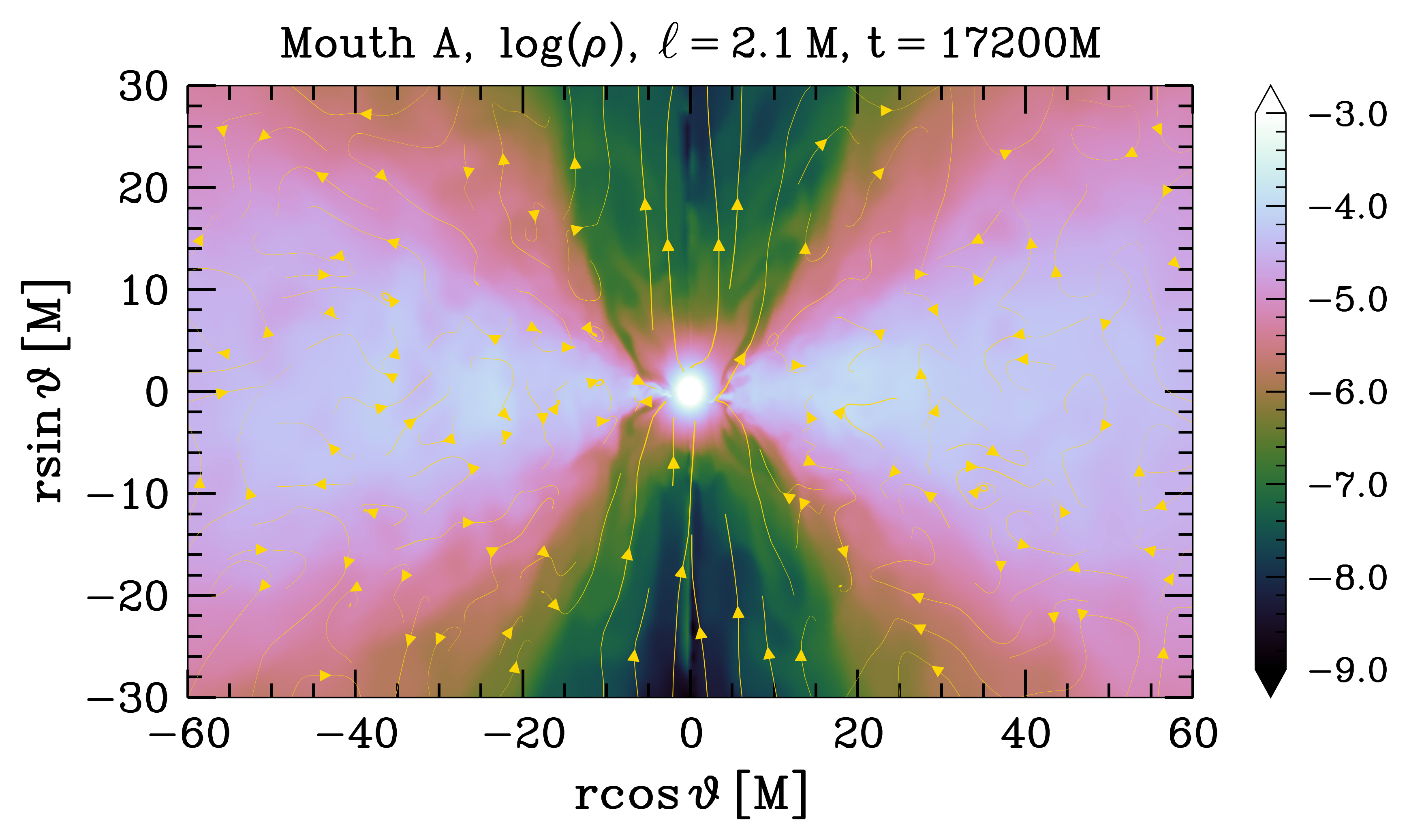}
    \includegraphics[width=0.49\linewidth]{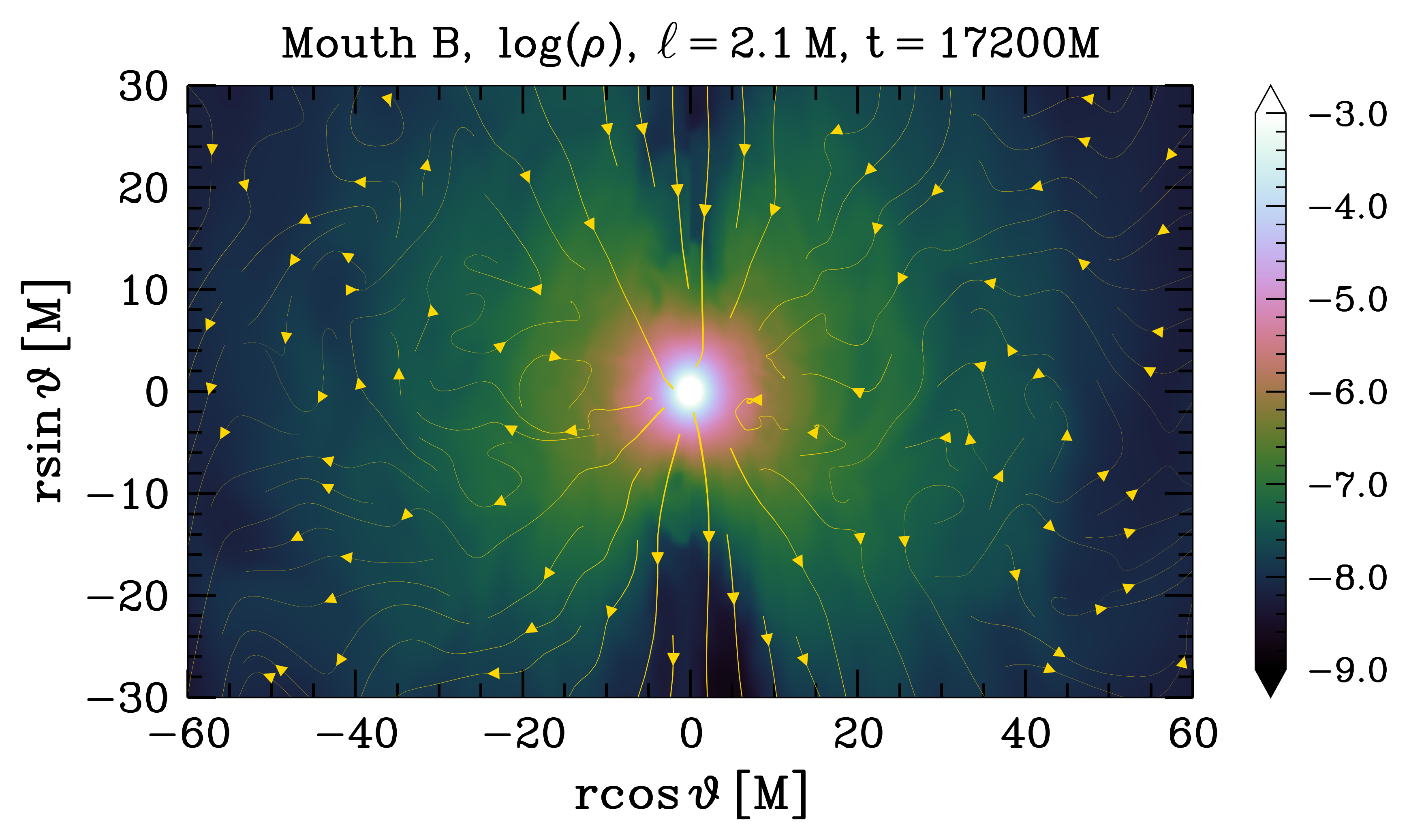}
    \caption{Top: $\theta$-averaged quantities at $t=20\times10^4M$ projected onto an embedding diagram of the wormhole for $\ell=2.1M$. The center of the throat exhibits a hot overdensity--- a wormhole cloud. Bottom: Meridional ($x=0$) rest-mass density plot at mouth A (left) and mouth B (right) with magnetic field lines.}
    \label{fig:embdens}
\end{figure*}

\section{Results}
\subsection{Accretion from disk to throat} 

The onset of turbulence by the MRI triggers accretion of an initially equilibrated torus onto mouth A (Fig. \ref{fig:qvst}). The gas initially flows freely through the throat ($r=0$) and emerges from mouth B. Because the infalling gas is bound on average ($-hu_t <1$, where $h$ is the specific enthalpy), most of it precipitates back into the throat. This fall-back material then shocks against the inflowing gas, drastically increasing the temperature and density in the throat by several orders of magnitude with respect to the initial atmosphere values, as shown in the mass evolution integrated over $|r|<5M$ around the throat (Fig. \ref{fig:qvst}). Since $r=0$ is a saddle point of the effective gravitational potential, the gas sloshing in the throat dissipates its kinetic energy and accumulates there. The surface-averaged rest-mass density across the throat  $\langle \rho \rangle : = {\int \rho d\mathcal{A} }/{\int d\mathcal{A}}$, with $d\mathcal{A}:=\sqrt{-g} d\theta d\phi$, grows $\sim 4$ orders of magnitude above that of the disk (Fig. \ref{fig:qvsr}). Similarly, the temperature in the throat increases by $\sim 3$ orders of magnitude with respect to the disk. In the top panel of Fig.  \ref{fig:embdens}, we show an embedding diagram of the wormhole for different $\theta-$averaged quantities illustrating how matter accumulates in the throat. 

\begin{figure}[ht!]
    \centering
    \includegraphics[width=\columnwidth]{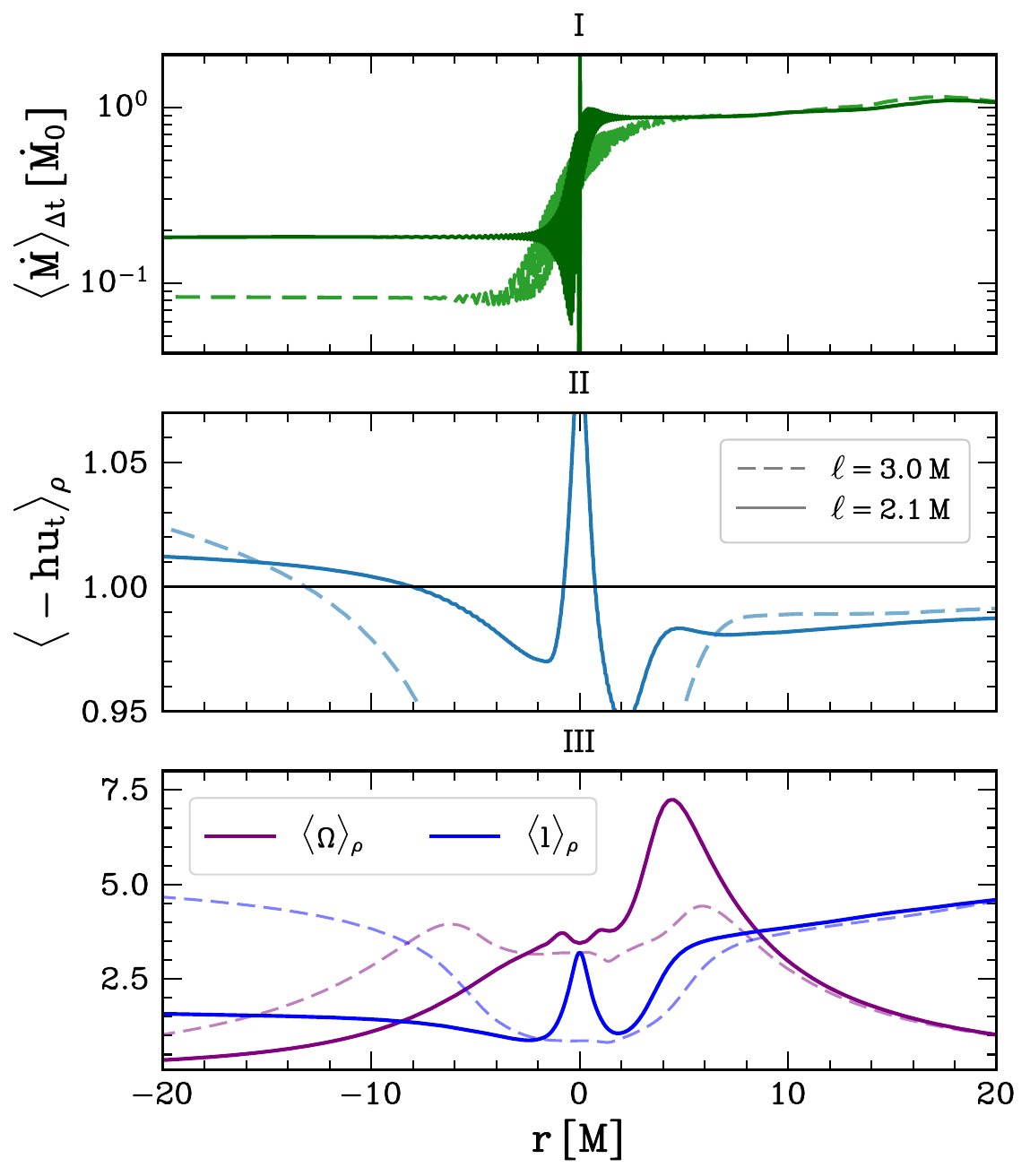}
    \caption{We plot different quantities for $\ell=2.1M$ in thick lines and $\ell=3M$ in dashed lines.
    (I): Time-averaged mass flux (accretion on $r>0$ and outflow on $r<0$) as a function of radius for simulations with $\ell=2.1M$ ($\dot{M}_0 = 2 \times 10^{-4}$) and $\ell=3M$ ($\dot{M}_0 = 0.75\times 10^{-4}$). 
    (II): Specific internal energy time-averaged.
    (III): Specific angular momentum (blue lines) and angular velocity (purple lines) as a function of radius, averaged on the sphere and in time between $t \in [10, 20]\times 10^{4} M$, weighted with rest-mass density}
    \label{fig:qenvsr}
\end{figure}

On mouth A, the over-density forming on the throat acts as a hard-surface inner boundary condition to the accretion disk,  restructuring the flow in the first $t\approx 10^4 M$; after this period, the mass flux, $\dot{M}:=-\int u^r \rho d\mathcal{A}$, on mouth A reaches a new inflow equilibrium (at $r<15M$) with an average value of $\langle \dot{M} \rangle_{A} \approx 2 \times 10^{-4}$ (Fig. \ref{fig:qenvsr}I) 
The mass flux on mouth B settles into an outflow steady-state with an average value of $\langle \dot{M} \rangle_{B} \approx 3\times 10^{-5}$. The imbalance in mass flux across mouths A and B indicates that part of the matter is deposited in the throat, consistent with the linear growth of mass (see Fig. \ref{fig:qvsr}).

\subsection{An engine for thermal winds} The over-density at the throat forms a hot, rotating bound structure extending to both sides with an average radius of $|r_{\rm cloud}| \approx 5 M$ as shown in Figs. \ref{fig:embdens} (bottom panels) and \ref{fig:qvsr}. This ``wormhole cloud'' is differentially rotating with an average angular velocity, $\langle \Omega \rangle_{\rho}:= \langle u^{\phi}/u^{t} \rangle_{\rho}$ peaking around $r \approx 5 M$, while decreasing towards the center (Fig. \ref{fig:qenvsr}III). Outside the cloud, $\langle \Omega \rangle_{\rho}$ follows a Keplerian profile $\propto r^{-3/2}$, whereas the interior obeys $d\langle \Omega \rangle_{\rho}/dr>0$ (a $d\langle \Omega \rangle_{\rho}/dr<0$ profile would decelerate the core and not be sustainable), similar to neutron stars remnants in binary neutron star mergers \cite{kastaun_black_2013}.  At the throat, $\Omega$ flattens and transitions to a sub-Keplerian profile on mouth B. The average specific angular momentum $\langle l \rangle_{\rho} := \langle  -u_{\phi}/u_t \rangle_{\rho}$ in mouth A follows a Keplerian profile in the bulk of the disk (Fig. \ref{fig:qenvsr}III), starts decreasing inside the ISCO ($r=5.2 M$), and peaks again at the center of the throat. This can be understood as the effect of a net positive torque exerted by the outer envelope of the cloud toward the center as consistent with $d\langle \Omega \rangle_{\rho}/dr<0$ in the interior.

\begin{figure}[ht!]
    \centering    \includegraphics[width=0.9\linewidth]{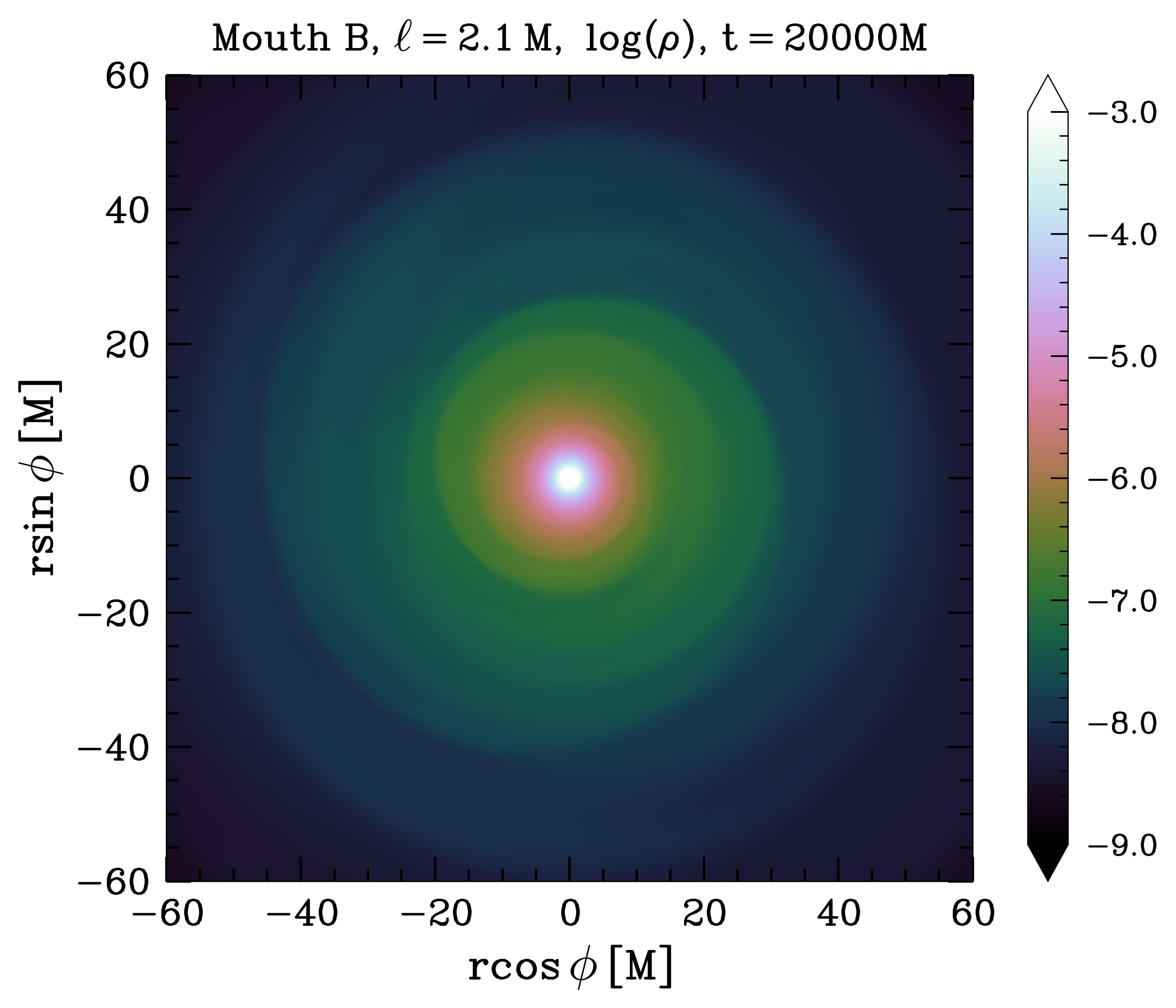}
    \caption{Equatorial ($z=0$) rest-mass density plot at mouth B for $\ell=2.1M$. The wind has $m=1$ density modes triggered by oscillations in the wormhole cloud.}
    \label{fig:eqplot}
\end{figure}

The cloud receives matter from the accretion disk in mouth A and launches a net outflow from mouth B. Because the flow coming from mouth A is originally bound on average and the spacetime is symmetric on both sides of the throat, a mechanism to release energy must be in place to power the outflow. Indeed, this extra energy comes from the rest-mass density deposited in the throat, shown in the imbalance of mass flux across the throat (Fig. \ref{fig:qenvsr}I). The wormhole cloud functions as an engine: matter deposited in the throat on mouth A is compressed, loses kinetic energy, and heats up the cloud. The rest-mass density in the throat increases, and a portion of the fluid in mouth B gains thermal energy in the process, producing a wind (Fig. \ref{fig:embdens}, bottom {right}). 

\subsection{Properties of the wormhole wind} The outflow on mouth B has an asymptotic velocity of $v_{\infty}  = \sqrt{1-1/(-hu_t)^2} \approx 0.2 c$ becoming unbound at {$r\approx -8\:M$} as shown in Fig. \ref{fig:qenvsr}II. Accreted material from mouth A passing through the throat increases its temperature and entropy due to shocks (see Fig. \ref{fig:embdens}, top panel, and Fig.\ref{fig:qvsr}); the entropic outflow on mouth B then cools down as it expands (Fig. \ref{fig:qvsr}). Non-axisymmetric density modes in the wormhole cloud endow this wind with an $m=1$ spiral pattern (Fig. \ref{fig:eqplot}). The wind on mouth B has a sub-Keplerian-specific angular momentum profile, which remains almost constant as the wind expands. The surface-averaged properties of the wormhole wind are described fairly well by a spherically-symmetric stationary wind (dotted-dashed lines in Fig. \ref{fig:qvsr}) that decays as $\rho_{\rm WS} \propto r^{-2}$ for large distances as the wind reaches a constant terminal velocity (see Appendix). The efficiency of the wormhole wind can be defined from the jump of mass flux (Fig. \ref{fig:qenvsr}I), as  $\eta := \langle \dot{M} \rangle_{B}/\langle \dot{M} \rangle_{A} \times 100 \sim 20\%$. 
. 

\subsection{Magnetic field in the throat} 
Magnetic fields are necessary to produce turbulent MRI-driven stresses in the disk, triggering accretion from mouth A and advecting magnetic flux into the throat. The magnetic field strength increases in the center of the throat as it is compressed by gravity, see upper left diagram in Fig. \ref{fig:embdens}. The poloidal magnetic flux (using Gaussian units), $\Phi:= (1/2) \int |B^r| d\Omega$, reaches a steady-state (Fig. \ref{fig:qvst}) with a mass-flux normalized average value of $\sqrt{4 \pi} \Phi/\sqrt{\dot{M}} \approx 15-20$ on mouth A. Part of this magnetic flux is trapped in the throat and part of it emerges on the other side; indeed, on mouth B we observe a magnetic flux that is on average a factor of $4-5$ less than the flux on mouth A. The spacetime is non-rotating and thus the system does not produce strong Poynting luminosities by a Blandford--Znajeck type process, which depends on $\Phi$. If the system has an ergosphere, e.g. as in the Teo wormhole metric \citep{teo1998rotating}, these results show that one side (mouth A) would likely produce a Poynting outflow --- due to magnetic reconnection in the current sheets \cite{East:2018ayf}--- with a larger luminosity than the other side.
The global topology of the field is conserved across $r=0$ but with an inverse polarity (Fig. \ref{fig:embdens}, bottom right). Winds launched from mouth B are weakly magnetized, with average values of $\sigma = (b^2/2)/\rho \approx 10^{-3}$, see Fig. \ref{fig:qvsr}.

\subsection{Comparison between different throat sizes}

In the simulation with throat's size $\ell = 3M$, a similar wind-emitting cloud is formed, albeit with some noticeable differences. The cloud is bigger for $\ell = 3M$ and the wind becomes marginally unbound further away, at a typical radius of $r\sim 14 M$ (see Fig. \ref{fig:qenvsr}II). The conversion of gravitational energy to kinetic/thermal energy is expected to be less efficient in a wormhole with a bigger throat because of the shallower effective potential. However, it is easier to unbind a fluid element that is trapped in the potential if the throat is bigger. We find that the efficiency for emitting winds in the simulation with $\ell=3M$ is actually smaller, $\eta \sim 5 \%$ (see Fig. \ref{fig:qenvsr}I).

\begin{figure}[ht!]
    \centering    \includegraphics[width=1.0\linewidth]{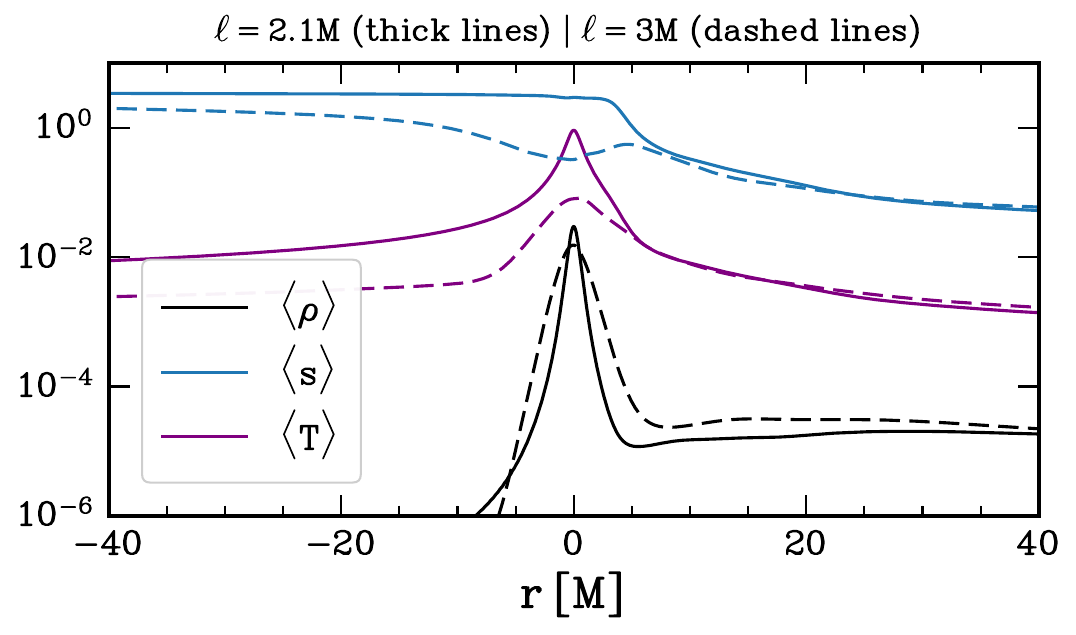}
    \caption{Surface-avareged rest-mass density, entropy, and temperature for $\ell=2.1$ (thick lines) and $\ell=3M$ (dashed lines).}
    \label{fig:l3l2comp}
\end{figure}

\begin{figure}[ht!]
    \centering    \includegraphics[width=1.0\linewidth]{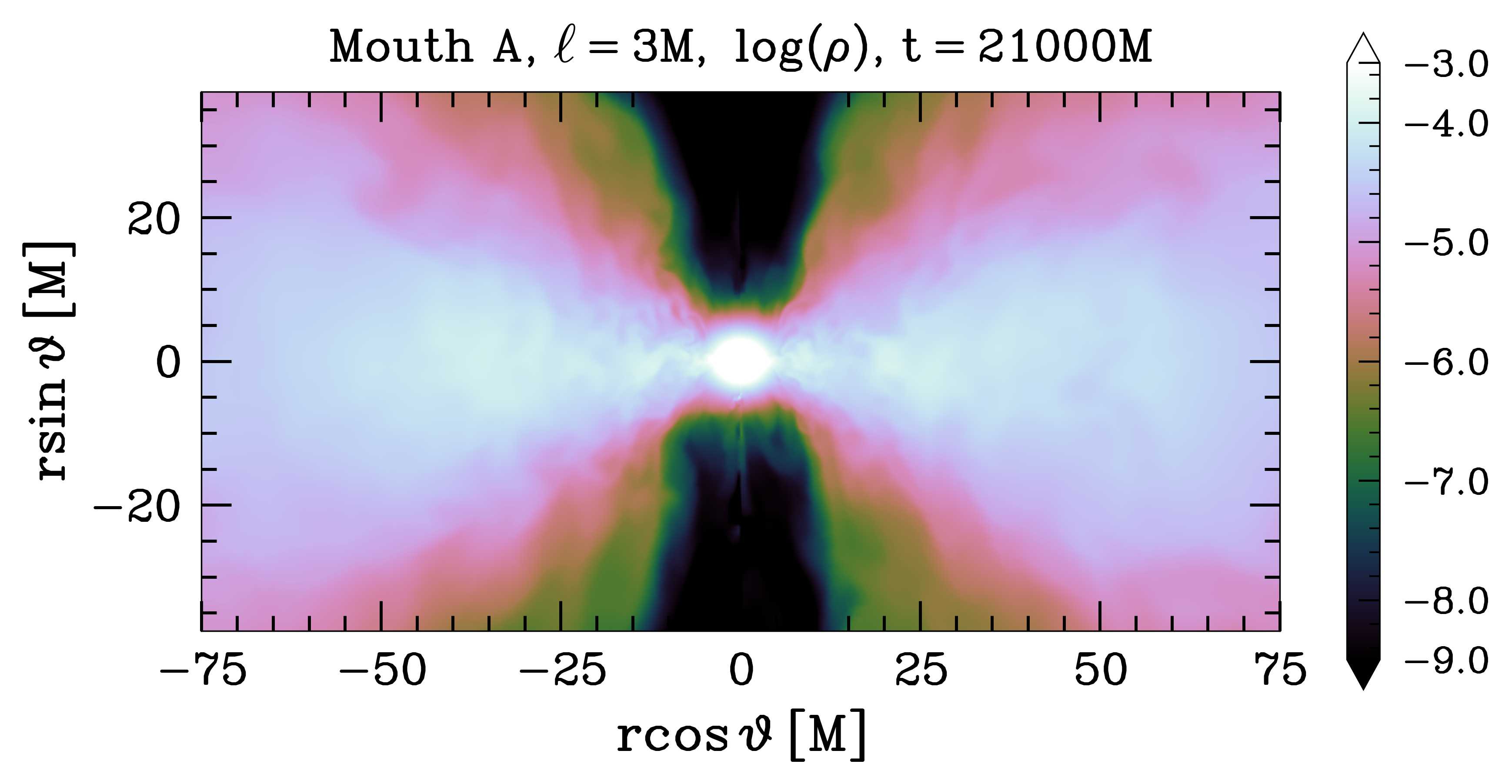}
    \includegraphics[width=1.0\linewidth]{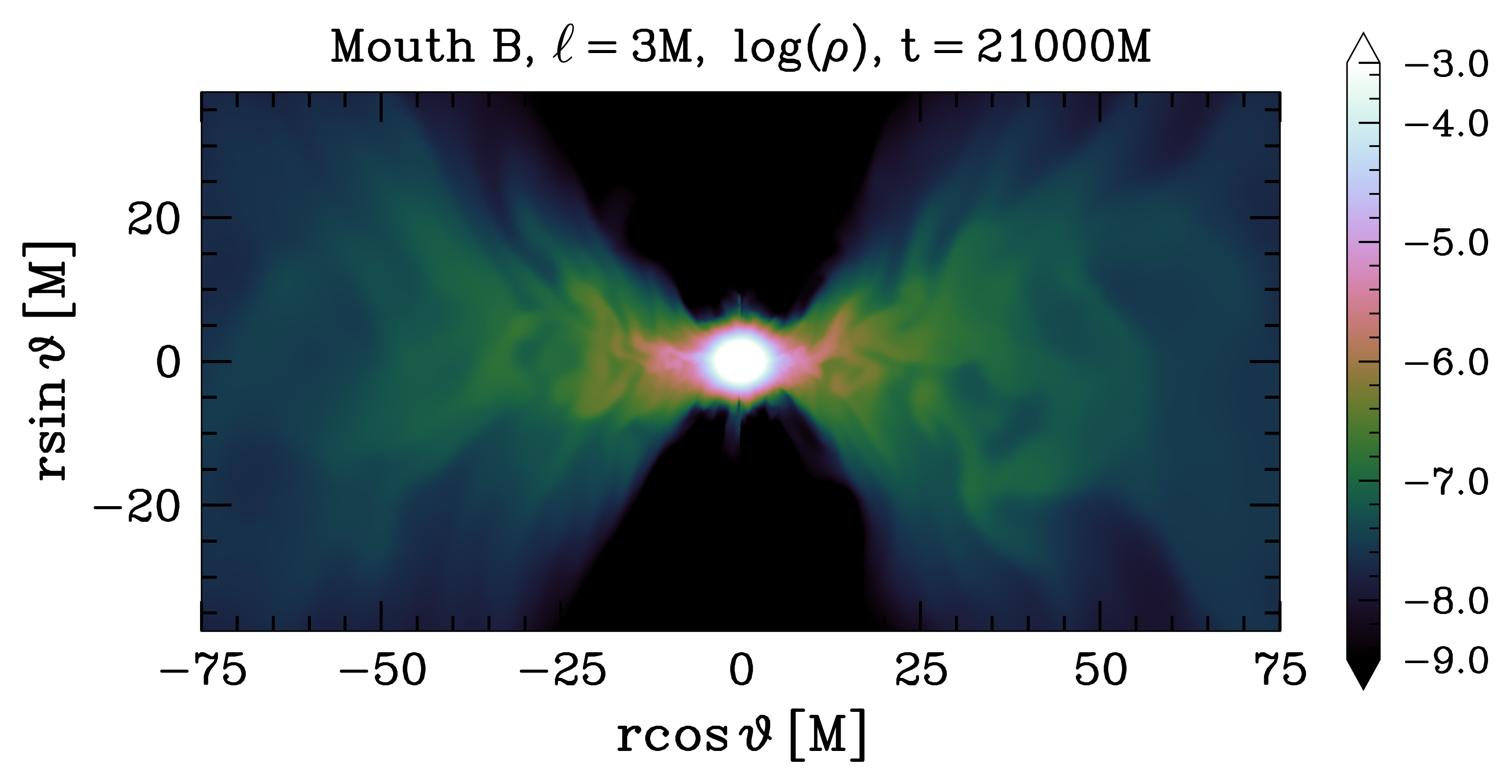}
    \caption{Meridional ($x=0$) rest-mass density plot at mouth A (top) and mouth B (bottom) for $\ell=3M$.}
    \label{fig:l3rho}
\end{figure}

{In $\ell=3M$, the flow does not form a strong shock on the throat and does not heat up as much, transitioning from mouth A to B smoothly, see Fig. \ref{fig:l3l2comp}. In $\ell=2.1M$, because of the strong heating across the throat, the wind is launched quasi-isotropically on mouth B (Fig. \ref{fig:embdens}, bottom right), while in $\ell=3\:M$ the wind is colder and propagates closer to the equatorial plane, i.e. with a smaller scale height, see Fig.\ref{fig:l3rho}.} 

Even though the outflow is less massive than in the $\ell=2.1M$ case, we find that asymptotic outflow velocities are larger for $\ell=3M$ with $v_{\infty} \sim 0.3c$, as shown in the specific internal energy in Fig. \ref{fig:qenvsr}.  This difference is explained by the larger rotational energy of the outflow in $\ell=3M$. Indeed, in the $\ell=2.1M$ case, matter with high specific angular momentum is trapped in the throat, while in the $\ell=3M$ case, most of the angular momentum is carried by the wind, as shown in Fig. \ref{fig:qenvsr}III, dashed lines. Notice that the rotational profile for $\ell=3M$ is symmetric with respect to the throat, contrary to $\ell=2.1M$. 

\section{Astrophysical consequences of accreting wormholes}
\rvw{
In our simulations, we assumed that the total mass of the accreting gas is negligible compared to the mass of the wormhole. The predicted linear growth of the density in the throat, however, implies that the gas can eventually become dynamically important, e.g. triggering the collapse of the wormhole into a BH. Converting code units to physical units (see Appendix) we find that $\rho_{\rm th} \sim 1$ gr cm$^{-3}$ $(\dot{M}/\dot{M}_{\rm Edd}) (M_{\odot}/M)$, and temperatures of $T_{\rm th} \sim 10^{11} K$ (which might be sufficient to ignite hydrogen fusion for stellar-mass systems). The mass of the gas in the throat then would evolve as 
\begin{equation}
M_{\rm th} \sim \frac{ 4\pi}{3} r_{\rm th}^3 \rho_{\rm th} \sim 10^{-6} M_{\odot} \frac{\dot{M}}{\dot{M}_{\rm Edd}} \frac{M_{\odot}}{M} \frac{t}{\rm years},    
\end{equation}
which can reach $M_{\rm th} \sim M$ in less than a Hubble time for stellar-mass systems. This implies that astrophysical wormholes might have a finite life span.}

\rvw{Different from an accreting BH, a wormhole does not have an absorbing boundary. Stationary accretion models on wormholes have focused on the properties of the disk emission near the ISCO \citep{bambi2021astrophysical, harko2008electromagnetic, harko2009thin} and ray-tracing assuming the light can pass through the throat \cite{bambhaniya2022thin}. They ignore, however, the self-interaction of the flow in the throat, which changes the global properties of the system and its EM emission. On mouth B, the initially optically thick outflow expands and eventually becomes optically thin, producing thermal emission.

In our accretion model, all heat is retained in the flow, and for high-accretion rates, the gas would be optically thick to radiation. As the wormhole wind on mouth B expands and adiabatically cools, the plasma will eventually become optically thin. Assuming a stationary mildly-relativistic wind, where $u^r = \Gamma v^r \approx 0.2$ as extracted from the simulation, the photosphere radius, $r_{\rm ph}$, is placed where the optical depth becomes one
\begin{equation}
    \tau = \int^{\infty}_{r_{\rm ph}} \kappa_{\rm e} \rho_{\rm B}(r) dr = 1,
\end{equation}
where $\rho_{\rm B}$ is the density at mouth B ($r<0$). We then find that
\begin{equation}
    r_{\rm ph} \sim \frac{\dot{M}_{\rm wind} \kappa_{\rm e}}{4 \pi v^r}  \approx 15 \: \frac{G M}{c^2} \times \Big(\frac{\dot{M}}{\dot{M}_{\rm edd}}\Big),
\end{equation}
where $\dot{M}_{\rm wind}= 0.25 \dot{M}$ is the mass-flux of the wind, $\dot{M}$ is the accretion rate onto mouth A, and $\kappa_{\rm e}$ is the electron scattering opacity. At the photosphere, we assume that the effective temperature of thermalized photons is set by the total thermal energy:
\begin{equation}
    u = \rho \frac{k_{\rm B}}{m_{\rm p}(\Gamma-1)} T_{\rm eff}+ a T_{\rm eff}^4,
\end{equation}
where $a$ is the radiation constant. We find that, at the photosphere, $T_{\rm eff} \sim  10^{7} K$, which peaks on the far UV-X-rays for stellar-mass systems; here we plugged the numerical value of $u$ and $\rho$ from our simulation, $M=M_{\odot}$, and $\dot{M}=\dot{M}_{\rm edd}$. 

The formation of a hot overdensity pouring on both sides of the throat will drastically change the electromagnetic radiation compared to an accreting BH. In particular, on mouth A ---the accreting side--- there will be no shadow because the cloud size is larger than the light ring, contrary to previous expectations \cite{akiyama2022first}.}

\section{Conclusions} 

We have performed the first 3D GRMHD simulations of accreting traversable wormholes. We found that the magnetized plasma accumulates in the throat and forms a rotating hot cloud which is $\sim 10^4$ denser than the disk. The wormhole cloud acts as an engine, taking gas from the accretion disk on one side, depositing part of it in the throat, and launching a mildly relativistic thermal wind on the other side with an efficiency of $\sim 20 \%$ for a throat of size $\ell = 2.1M$. \rvw{These results entail novel predictions for its EM radiation: the hot rotating cloud in the throat will shine beyond the light ring of the spacetime and thus there will be no shadow, even if the outer metric is similar to a BH. We estimate that the cloud wind will produce thermal emission peaking in the UV/X-rays for stellar mass systems. Moreover, we show that the hot cloud grows linearly in time and could eventually trigger the collapse of the wormhole into a BH.}

\rvw{Accumulation of gas in the throat occurs because the throat is a minimum stable point in the gravitational potential, a feature in most traversable wormholes; we then expect our main predictions to qualitatively hold in other realizations of spherically symmetric wormholes. We notice, however, that the wind efficiency depends on the generated heat in the throat; for accretion flows that are able to cool efficiently, matter will clump in the throat and the wind could be less powerful. Different initial conditions and the inclusion of spin might alter some aspects of the system and can be studied with the techniques introduced here (c.f. Refs.\cite{kocherlakota2023toward,chatterjee2023energy} for GRMHDs onto other exotic objects). Our work shows that simulations are crucial to producing realistic models of accreting wormholes}\\

\acknowledgments


We thank Daniel Siegel for reading the draft and providing insightful comments. This research was enabled in part by support provided by SciNet (www.scinethpc.ca) and Compute Canada (www.computecanada.ca). LC is a CITA National fellow and acknowledges the support of the Natural Sciences and Engineering Research Council of Canada (NSERC), funding reference DIS-2022-568580. 
H. Y. is supported by the Natural Sciences and
Engineering Research Council of Canada and in part by
Perimeter Institute for Theoretical Physics.
Research at Perimeter Institute is supported in part by the Government of Canada through the Department of Innovation, Science and Economic Development Canada and by the Province of Ontario through the Ministry of Colleges and Universities. 
G.E.R. acknowledges financial support from the State Agency for Research of the Spanish Ministry of Science and Innovation
under grants PID2019-105510GB-C31AEI/ 10.13039/501100011033/ and
PID2022-136828NB-C41/AEI10.13039/501100011033, and by ``ERDF: A way of
making Europe'', by the "European Union", and through the ``Unit of Excellence Mar\'ia de Maeztu 2020-2023" award to the Institute of
Cosmos Sciences (CEX2019-000918-M). Additional support came from PIP 0554 (CONICET).

\section{Appendix}

\subsection{Properties of the Simpson--Visser wormhole}

The spacetime of a spherically symmetric wormhole can be characterized by its redshift function, $\varphi(r)$, and shape form $b(r)$ \cite{morris1988wormholes}. The redshift factor determines the acceleration of the stationary observer at a fixed $r$ as well as the time of flight from nearby regions as seen by an asymptotic observer; the shape factor $b(r)$, determines the embedding diagram of the wormhole. The two functions set the tidal deformations in the throat which will affect the fluid.

We can analyze time-like geodesics in this wormhole spacetime taking into account the motion equation for the radial velocity:
\begin{equation}
    E = \dot{r}^2 + V_{\rm eff}(r)
\end{equation}
where $E$ is the conserved energy and the effective potential is:
\begin{equation}
    V_{\rm eff}(r) : = \Big(1-\frac{2\:M}{\sqrt{r^2+\ell^2}} \Big) \Big( \frac{L^2}{(r^2 + \ell^2)}+1 \Big)
\end{equation}
\begin{equation}
    \equiv V^{\rm Schw}_{\rm eff}(r) + \frac{(M/ \ell)}{(r/\ell)^3} + \mathcal{O}(r^{-4}).
\end{equation}

\begin{figure}[ht!]
    \centering
    \includegraphics[width=\columnwidth]{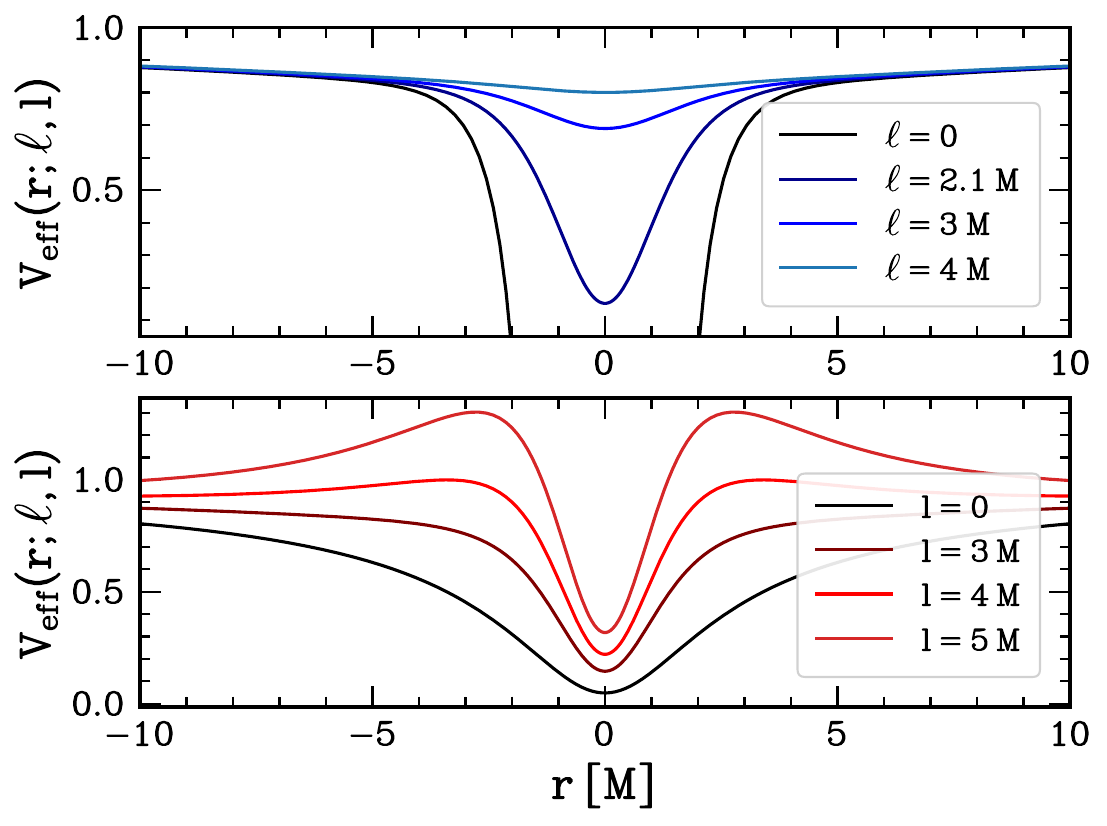}
    \caption{Top: effective potential for fixed specific angular momentum $l=3M$ and different throat sizes. Bottom: effective potential for fixed throat size $\ell=2.1M$ and different values of specific angular momentum.  }
    \label{fig:effvsr}
\end{figure}

From this effective potential, we can immediately see that the particle motion is very similar to a Schwarzschild BH outside the ISCO. A remarkable feature of the wormhole spacetime is that there is a valley in the potential at $r=0$. Let us consider the zero specific angular momentum case. In a black hole, a particle released at rest from a finite distance falls directly into the horizon. In the case of the wormhole, a particle that is released at a finite radius $r_0$ will fall into the throat, emerge from the other side and go back, oscillating between $\pm r_{0}$. If a particle is released from infinity, it will reach infinity on the other side if there is no dissipation of energy; if the particle loses energy in the process, then it will get easily trapped in the throat oscillating back and forth until it settles in $r=0$ where it can rest. 

To eject matter to the other side, the fluid must acquire some additional kinetic energy from MHD processes or thermal energy. In our simulations, the incoming fluid dissipates kinetic into thermal energy, which is transferred to an outer portion of the fluid that will escape. If the throat is smaller (smaller $\ell$), the effective gravitational potential is deeper and the gas will be able to gain and dissipate more kinetic energy from the gravitational field, launching a more efficient wind. On the other hand, a smaller throat also means that the gas is more trapped in the gravitational field, acting against the wind launching. However, because gas thermal pressure increases the size of the cloud, the outer parts of the bound structure will gain sufficient energy to escape as they receive thermal energy from the interior.

\subsection{Embedded diagram}

We construct the embedding diagram following Refs.  \cite{morris1988wormholes}. We embed a 2D spatial sheet of the wormhole metric fixing $\theta = \pi/2$ and $t=0$, with line element $ds^2 = f(t) dr^2 + (r^2+\ell^2)d\theta$, into an Euclidean 3-dimensional spacetime with metric $ds_{E^3}^2 = dz^2 + dR^2 + R^2 d\varphi$ in cylindrical coordinates. Assuming a surface $z=z(r)$ and $R=R(r)$, we have:
\begin{equation}
    ds_{E^3}^2 = \Big(\Big[\frac{dR}{dr}\Big]^2 + \Big[\frac{dz}{dr}\Big]^2\Big)dr^2 + R(r)^2 d\varphi.
\end{equation}

Imposing $ds^2 = ds_{E^3}^2$, we have $R(r)^2 = (r^2+\ell^2)$ and  the following equation for $z$:
\begin{equation}
\frac{dz}{dr} = \sqrt{\frac{1}{f(r)}- \frac{r^2}{r^2+\ell^2}} 
\end{equation}
which can be solved numerically to give a surface $z(r)$.

\subsection{Spherically-symmetric wind solution for wormholes}

Let us consider a radial and stationary ideal fluid with four-velocity $u^a = u^t(r) \partial_t + u^r(r)\partial_r$ in our wormhole spacetime 
(Eq. \ref{eq:metric}). Considering the constraint  $u^a u_b = -1$, together with the ideal gas equation of state $p=(\Gamma-1) u$, our unknown fluid variables are given by  $u^r(r)$,  the enthalpy, $h(r)= 1 + (u+p)/\rho \equiv 1 + \Gamma u/\rho$, and the rest-mass density $\rho(r)$. We thus need three equations to solve the system \cite{bahamonde2015accretion}. 

We start using the mass continuity equation, $\nabla_{a}(\rho u^a) =0$, obtaining
\begin{equation}
    4 \pi \rho u^{r} (r^2 + \ell^2) = \dot{M}.
\end{equation}

A basic feature of a wind solution that has reached a terminal constant velocity $u^r=u_{\infty}$ is that the density decays as $\rho \propto r^{-2}$ at large separations. In a BH spacetime, the density necessarily diverges as it approaches the horizon; in the case of the wormhole, the density and the velocity can remain finite in the throat. 

We now consider the conservation equation $\nabla_a T^{ab} =0$, which can be separated into a momentum equation (projected to $h_{ab}$) and an energy equation (projected to $u^a$).  Since the fluid and spacetime are spherically symmetric and stationary, the momentum equation is equivalent to the Bernoulli equation:
\begin{equation}
   - h u_t =  -h \sqrt{(u^r)^2 + f(r)} = E.
\end{equation}

The remaining equation is the energy equation $u^a \nabla_{a} T^{ab} = 0$, which can be rewritten as the first law of thermodynamics for an adiabatic fluid:
\begin{equation}
   \frac{d ( \rho + u) }{d\rho} = h
\end{equation}
which gives $u=K \rho^{\Gamma}$. With these equations, we find a solution in terms of $\dot{M}$, $E$, and K. For a wind solution, we set $\rho_\infty = 0$, and thus $h_{\infty} =1$ and $E = \sqrt{(u^r_{\infty})^2+1}$. On the other hand $u^r(r) = \sqrt{E^2/h(r)^2 - f(r)}$, and then: 
\begin{equation}
  4 \pi  \rho(r) (r^2 + \ell^2) \sqrt{\frac{E^2}{1+\Gamma K  \rho(r)^{\Gamma-1}} - f(r)}  = \dot{M}.
\end{equation}

We can analyze the solution in terms of its value at the critical point when the velocity is equal to the sound speed; taking the differentials from the Bernoulli and mass conservation equations, we have:
\begin{equation*}
    \frac{du^r}{u^r} \Big(c_{\rm s}^2 - \frac{(u^r)^2}{(u^r)^2 + f(r)}\Big)+
\end{equation*}
\begin{equation*}
    \frac{dr}{r} \Big( 2 r^2 \frac{c_{\rm s}^2}{r^2+\ell^2} - \frac{r f'(r)}{2((u^r)^2 + f(r))}\Big) = 0,
\end{equation*}
where we define the relativistic sound speed $c_{\rm s}^2 =(\rho/h) \partial h/\partial \rho $. If the solution reaches the critical point at some radius $r_{\rm c}$ where $v^r_{\rm c} = u^r_{\rm c}/((u^r_{\rm c})^2+f(r_{\rm c})) = c_{\rm s, c}$, then to maintain regularity, the other parenthesis in the previous equation should be zero, i.e.:
\begin{equation}
    2 \frac{r_{c}^2}{r_{c}^2 + \ell^2} c_{\rm s, c} = r_{\rm c} \frac{f'(r_{\rm c})}{u_t^2},
\end{equation}
which means 
\begin{equation}
    u^r_{\rm c} = \frac{f'(r_{\rm c}) (r_{\rm c}^2 +\ell^2)}{4 r_{\rm c}} = \frac{M}{2\sqrt{r_{\rm c}^2 + \ell^2}},
\end{equation}
and
\begin{equation}
    c_{\rm s, c} = \frac{u^r_{\rm c}}{1- 3 u^r_{\rm c}}.
\end{equation}

With these equations, we can check if the solution reaches or not a critical point. It is interesting to check what happens if the critical point is on the throat ($r=0$). In that case, $c_{\rm s, c} = M/(2\ell -3 M) < 1$, while $u^r_{\rm c} = M/(2\ell) < 1/4$. In turn, this implies that the density in the throat must be $\rho(0) = (\dot{M}/M)/(2 \pi \ell)$. We leave a complete analysis of the wind properties of a wormhole for future work.

\subsection{Dependence on resolution}

In this section, we show that the bulk properties of gas in the simulation are robust under different resolutions and predict similar behavior for the inflow-outflow balance across the throat. We investigate three different resolutions: $180\times96\times96$ (LR), $224\times112\times112$ (MR), and $480\times128\times128$ (HR) for radial, poloidal, and azimuthal cells. We evolve the same wormhole accretion disk presented in the main text with $\ell =3M$ for each resolution for $t\sim10^4M$. 

\begin{figure}[ht!]
    \centering    \includegraphics[width=1.0\linewidth]{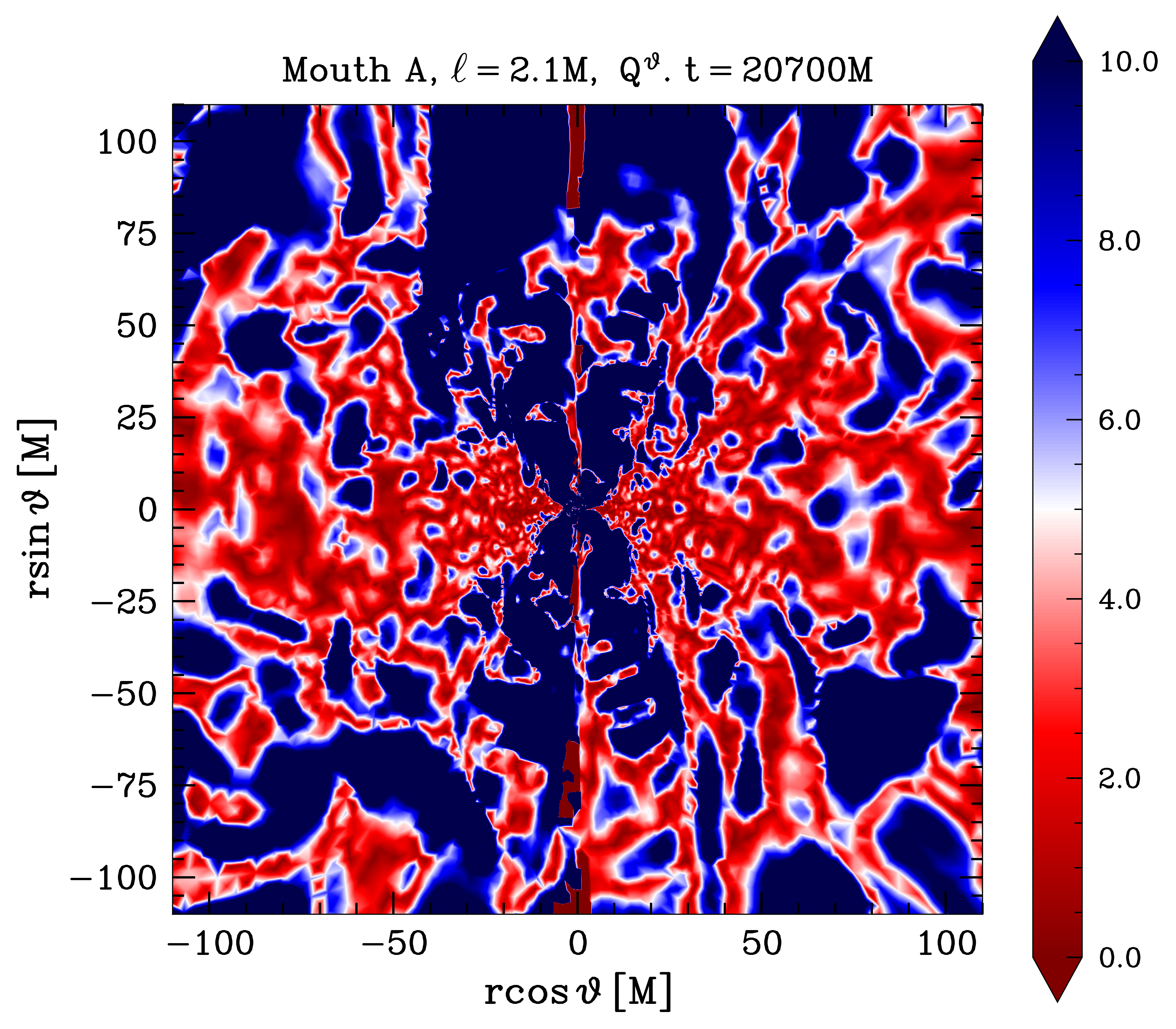}
    \caption{Meridional cut of the MRI quality factor for the $\theta$ direction for the $\ell \sim 2.1$ fiducial run.}
    \label{fig:qmri}
\end{figure}

The quality of the resolution to produce and maintain a turbulent state can be determined by the MRI quality factor, $Q$. Given the fastest growing wavelength of the linear MRI defined as 
\begin{equation}
    \lambda^{r,\theta} = \frac{2 \pi}{\Omega\sqrt{\rho h + b^2} } b^{r,\theta}.
\end{equation}
the quality factor is defined by the number of cells resolving this scale, $Q^{r,\theta} = \lambda^{r,\theta}/dx^{r,\theta}$. 
In our HR runs, the quality factor in different directions is an average
$Q^{r,\theta} \sim 10-15$, which is a good number to maintain a turbulent state, as discussed in Ref. \cite{porth2019event}. In Figure \ref{fig:qmri} we plot a snapshot of $Q^{\theta}$ for our fiducial run for $\ell=2.1$. Our LR run resolves only marginally the MRI $Q\sim 5$ but it is enough to sustain accretion through the duration of the simulation. 

\begin{figure}[ht!]
    \centering    \includegraphics[width=1.0\linewidth]{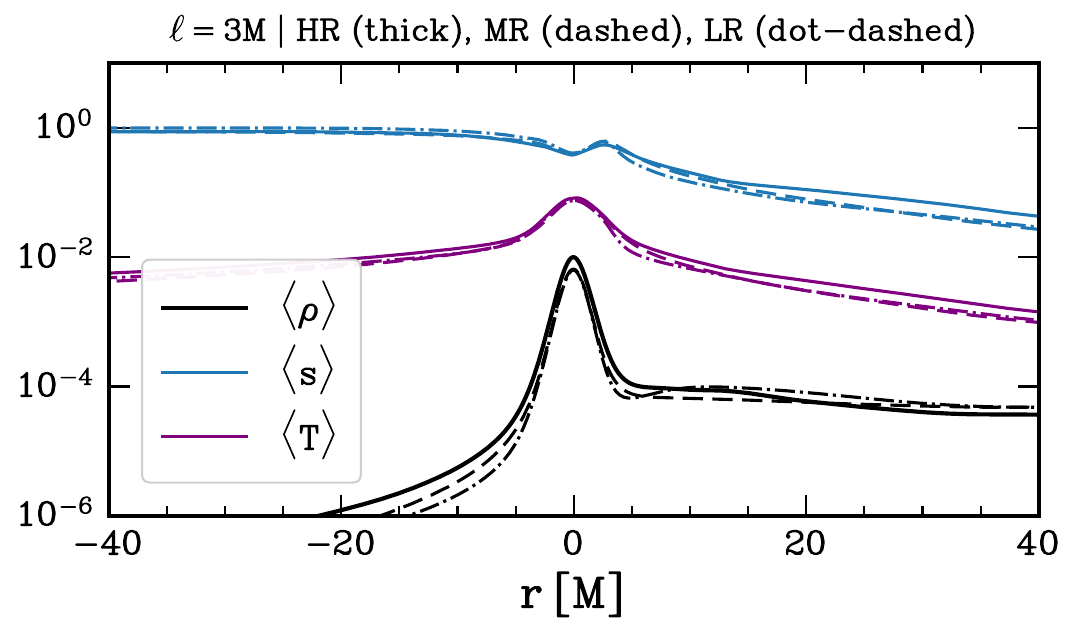}
    \caption{Surface-avareged rest-mass density, entropy, and temperature for $\ell=3M$ as a function of radius for different resolutions.}
    \label{fig:convqvsr}
\end{figure}

\begin{figure}[ht!]
    \centering    \includegraphics[width=1.0\linewidth]{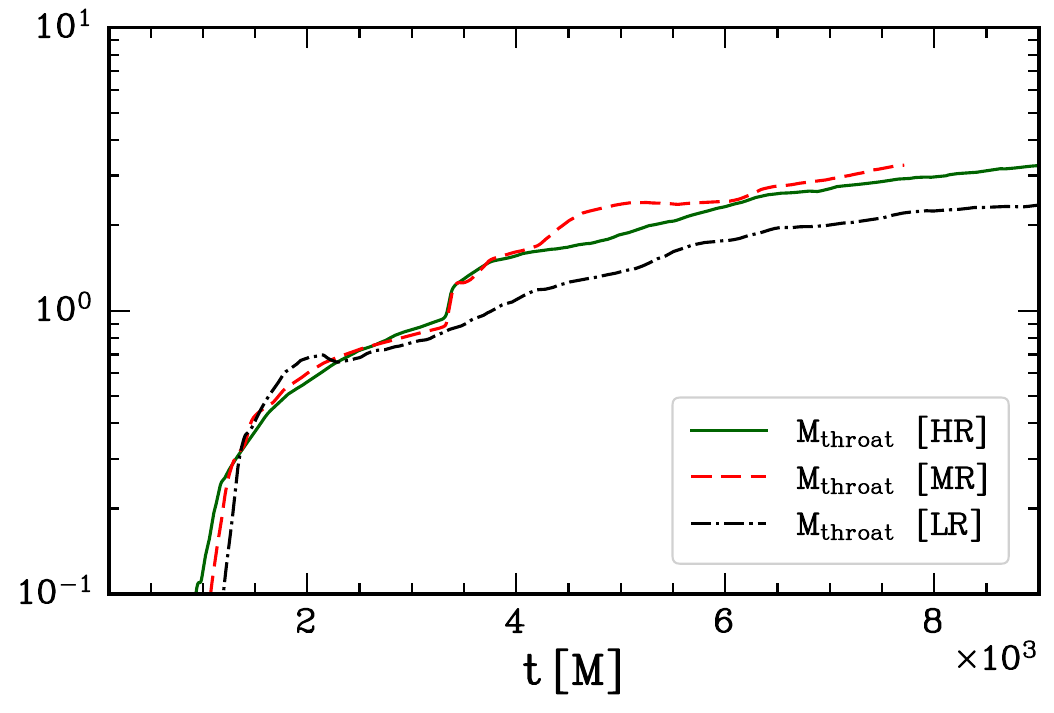}
    \caption{Time evolution of the mass contained in the throat for $\ell=3M$ at different resolutions.}
    \label{fig:convqvst}
\end{figure}

The time-averaged properties of the flow across the throat change only slightly with resolution, as shown in Figure \ref{fig:convqvsr}. In Figure. \ref{fig:convqvst}, we plot the mass on the throat as a function of time for each resolution. The mass flux is a bit higher for higher resolution, likely due to a better-resolved MRI in the accretion flow. 


\subsection{Transformation from code units to physical units}

Because the simulations are scale-free, the physical density is determined by choosing the mass of the spacetime, $M$, and the physical accretion rate, $\dot{M}$. These can be related to the quantities in code units using the conservation equation. Although we can choose any value for these two parameters, our thick-disk simulations are consistent with high or low accretion rates (with respect to the Eddington rate), which result in advection-dominated flows. Given the density in code units, $\rho_{\rm CU}$, and the accretion rate in code units, $\dot{M}_{\rm CU}$, taken from the simulation as a time average and radial average in $r\in [0,20M]$), the physical density is 
\begin{equation*}
    \rho = \frac{\rho_{\rm CU}}{\dot{M}_{\rm CU}} \frac{\dot{M}}{c} \Big(\frac{GM}{c^2}\Big)^{-2}
\end{equation*}
\begin{equation}
    =   2 \times 10^{-3} \textrm{gr  cm}^{-3} \frac{\rho_{\rm CU}}{\dot{M}_{\rm CU}} \Big(\frac{\dot{M}}{\dot{M}_{\rm edd}}\Big)
    \Big(\frac{M_{\odot}}{M}\Big),
\end{equation}
and the internal energy is given by:
\begin{equation}
    u = \frac{\rho_{\rm CU} T_{\rm CU}}{(\Gamma-1)\dot{M}_{\rm CU}} c^2 \frac{\dot{M}}{c} \Big(\frac{GM}{c^2}\Big)^{-2},
\end{equation}
\begin{equation}
    = 2 \times 10^{18} \textrm{erg cm}^{-3}  \frac{\rho_{\rm CU} T_{\rm CU}}{(\Gamma-1)\dot{M}_{\rm CU}} \Big(\frac{\dot{M}}{\dot{M}_{\rm edd}}\Big)
    \Big(\frac{M_{\odot}}{M}\Big)
\end{equation}
Because we have not included any type of cooling, the scale of the gas temperature is determined by the composition of the material, which we assume to be electrons. The physical temperature is then calculated from $\rho_{\rm CU}$ and $u_{\rm CU}$ through the equation of state as:
\begin{equation}
    T = c^2 m_{\rm p} k_{\rm B}^{-1} T_{\rm CU} = (\Gamma-1) \frac{u_{\rm CU}}{\rho_{\rm CU}} \times 10^{12} K,
\end{equation}
where $c$ is the speed of light, $m_{\rm p}$is the mass of the proton, and $k_{\rm B}$ is the Boltzmann constant.



\bibliographystyle{apsrev4-2}
\bibliography{ms}

\end{document}